\documentclass[aps,  twocolumn, showpacs, tightenline]{revtex4}

\usepackage{graphicx} 
\usepackage{epstopdf} 
\usepackage{epsfig}

\begin{document}

\title{Free-running InGaAs/InP Avalanche Photodiode with Active Quenching \\ for Single Photon Counting at Telecom Wavelengths}
\author{ R.~T.~Thew$^1$}\email{robert.thew@physics.unige.ch}
\author{ D.~Stucki$^1$}
\author{J-D.~Gautier$^1$}
\author{A.~Rochas$^2$}
\author{H.~Zbinden$^1$}

\affiliation{$^1$Group of Applied Physics, University of Geneva, 1211 Geneva 4, Switzerland}

\affiliation{$^2$ id Quantique SA, Chemin de la Marbrerie 3, CH-1227 Geneva, Switzerland}

\date{\today}

\begin{abstract}
We present an InGaAs/InP avalanche photodiode with an active quenching circuit on an ASIC (application specific integrated circuit) that is capable of  operating in both gated and free-running modes. The 1.6\,mm$^2$ ASIC chip is fabricated using CMOS (complementary metal oxide semiconductor) technology guaranteeing long-term stability, reliability and compactness. In the free-running mode we find a single photon detection efficiency of 10\,\% with  $<$\,2kHz of noise.
\end{abstract}

\pacs{85.60.Gz, 03.67.Dd, 42.68.Wt, 07.60.Vg}

\maketitle

\newpage


Single photon detection at telecom wavelengths has been increasing in importance since the mid 1990s as a range of applications have matured. These include quantum key distribution \cite{Gisin02a} and quantum optics in general, optical time domain reflectometry,  testing integrated circuits and eye-safe laser ranging. The most common detectors at these wavelengths are based on InGaAs/InP avalanche photodiodes (APDs). However, it has been widely accepted that these need to be used in a "gated mode" due to a variety of problems arising from afterpulsing and high noise levels \cite{Hiskett00a, Stucki2001a}. In gated mode, the diode is periodically reverse biased above the breakdown voltage and once a detection is made a long dead-time is applied,  effectively switching off the detector to avoid saturation from after-pulse effects. This dead-time can be considerably longer than the inverse of the gating frequency and is typically of the order of more than 10\,$\mu$s. The maximum detection frequency is then limited by the inverse of the detector's dead-time.

Alternative detection schemes have been studied. Germanium APDs were initially tested \cite{Owens94a,Lacaita94a} but whilst passive operation was possible, these required liquid nitrogen cooling and are no longer available. Near-IR (-infrared) PMT-MCPs (photomultiplier tube - microchannel plates) based on InP/InGaAs or InP/InGaAsP photocathodes are commercially available but suffer from a poor single photon detection probability, typically less than 1\,\%. More recently, superconducting materials have been proposed and tested \cite{Goltsman01a}, though they require cooling to 4\,K or lower and remain impractical for most applications. Schemes based on non-linear sum-frequency generation and silicon (Si) APDs have also been investigated, and although promising timing resolution and detection probabilities have been obtained, the detection scheme still suffers from  large dark count rates \cite{Langrock05a,Thew06a}.

There have been previous attempts at improving the operation of InGaAs/InP APDs, primarily studying different gating approaches to reduce afterpulsing and increase achievable gating rates \cite{Yoshizawa04a, Namekata06a}. Free-running operation of InGaAs APDs has previously been carried out \cite{Rarity00a} using a passive quenching approach with limited success. More recently, it has been suggested that when the diode is operated in asynchronous mode or in gated mode (with a gate duration typically longer than 10ns), an active quenching circuit must be used \cite{Cova04a}. However, recent work on active quenching circuits has focused on improvements in timing response (in Si APDs) \cite{Gallivanoni06a}.  In this Letter we show that using our active quenching ASIC (application specific integrated circuit), we are now able to obtain efficiency and noise characteristics previously only attainable with operation in a gated regime.  We first describe the active quenching ASIC operation before comparing the gated and free-running modes using the same APD. 


In general, to perform active quenching, a remote circuit senses the avalanche current pulse and then quickly reduces the voltage below breakdown to stop the impact ionization process. The operating voltage is subsequently restored after a certain dead-time. Ideally, this feedback loop duration must remain as short as possible in order to limit the number of carriers that flow into the diode during a gating pulse, limiting the number of trapped charges and thus the after-pulsing rate. InGaAs/InP APDs are generally cooled with a multi-stage Peltier element and combined with a quenching circuit made of discrete components. As a result of the form factor and power consumption considerations, the electronic circuit usually cannot be included in the cooled enclosure where the APD is located. This causes an increase in the quenching time which degrades the detector performance. Furthermore, as the APD must be connected to the circuit by cables, impedance mismatch and signal reflections can occur such that the applied voltage during the gate can be non-uniform, leading to variations of the single photon detection probability and timing resolution when the detector is switched on.  

\begin{figure}
\begin{center}
\epsfig{figure=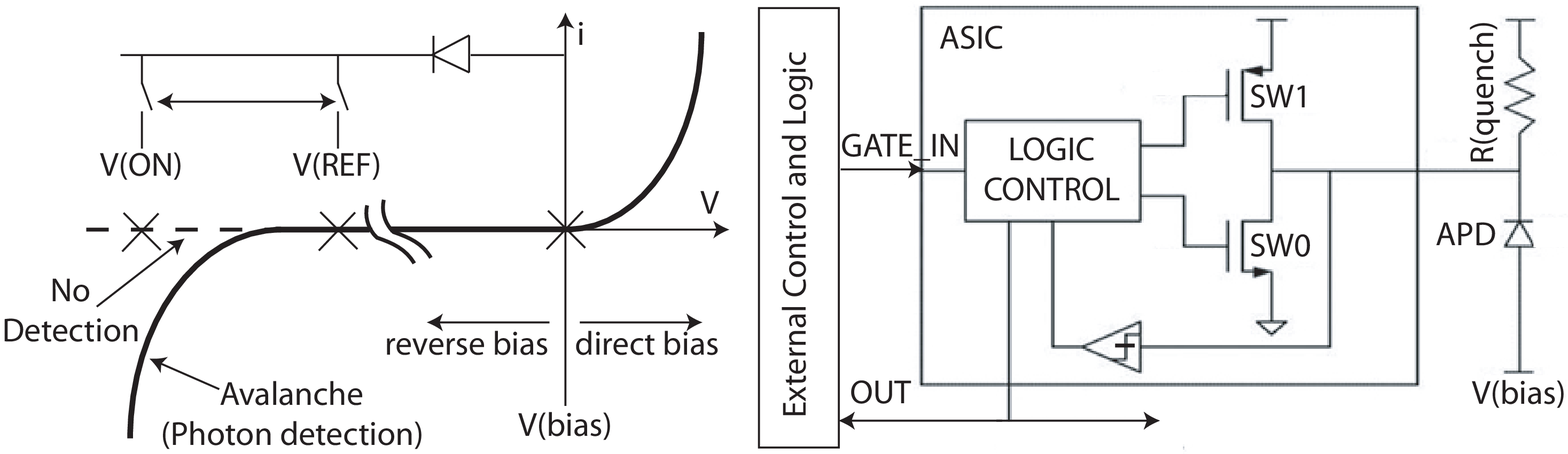,width=80mm,height=25mm}
\caption{(Left) Voltage-Current diagram for an APD. (Right) Simplified schematic of the electronic control and active quenching circuit and its connection to the APD.}\label{fig:circuit}
\label{fig:circuit}
\end{center}
\end{figure}
An ASIC has the advantage that it can be placed in close proximity to the APD, thus allowing for high-speed gating or free-running operation. This is important, as in this case the APD-ASIC separation is only 5\,mm, allowing us to minimise the parasitic capacitance. According to simulations, a quenching time of less than 5\,ns is expected. The circuit has been developed to function in either a gated or free-running mode  and the principle of its operation is illustrated in Fig.\,\ref{fig:circuit}. On the left we see an I-V curve where we have highlighted the bias voltage and a reference voltage V(REF) which holds the diode just below the breakdown voltage. When the detector is activated, or gated, the voltage is increased to V(ON) until it detects a photon and creates an avalanche, at which point it is actively quenched and the voltage is returned to V(REF). 

On the right of  Fig.\,\ref{fig:circuit}, there are two key components: an external control that governs free running operation; and a logic control that is on the ASIC chip which controls the gating and quenching of the APD. In the logic control, there is a level shifter that converts the GATE\_\,IN signal into pulses that control p-mos (SW1) and n-mos (SW0) switches. A short duration charge pulse is provided to SW1 to load the APD cathode to V(ON). Prior to SW1 closing, the SW0 switch is opened and it remains open until the end of the gate or an avalanche. Once an  avalanche is detected, a feedback signal is sent to the internal control logic and the SW0 switch is then closed, quenching the avalanche. More technical detail concerning this ASIC operation can be found in \cite{Rochas07a}, where some of us have recently demonstrated a notable reduction of the after-pulsing in gated mode operation. 

\begin{figure}
\begin{center}
\epsfig{figure=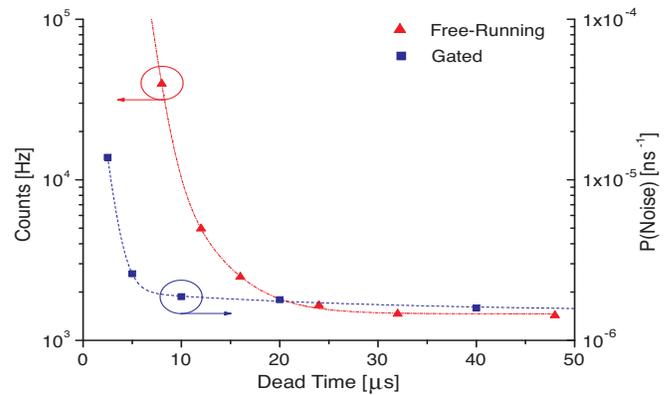,width=90mm,height=55mm}
\caption{Noise vs dead-time demonstrating the effect of after-pulsing in gated and free-running modes.}
\label{fig:afterpulses}
\end{center}
\end{figure}

This motivates us to begin by looking at the noise and after-pulsing characteristics.  In Fig.\,\ref{fig:afterpulses} we see a comparison between the noise for the system operating in both gated and free-running modes for different dead-times. This is done by observing the noise, for a given efficiency, as a function of the dead-time. In the gated regime the dead-time is varied by changing the triggering frequency $f_{Trig}$ (input at GATE\_IN in Fig.\,\ref{fig:circuit}) such that $\tau_{d} = 2\,/f_{Trig}$. The factor of two is due to the need for a reset signal, here given by the subsequent trigger pulse, for the logic control.  In the free-running regime the dead-time is set by the external control circuit. We initially set the temperature at 223\,K and the bias voltage for a detection efficiency of 10\,\%.

We see that, for the gated case, the effects of after-pulsing are negligible after 10\,$\mu$s, whereas in the free-running case, a longer dead-time is required $>$ 20\,$\mu$s. We can also vary the temperature and dead-time, depending on what detector characteristics we want. Recall that cooling (heating) the detector we will reduce (increase) the dark count level but increase (decrease) the after-pulse contributions, allowing the dead-time, and hence, the maximum detection rate, to be varied. The optimisation of these characteristics and a more detailed analysis of the dark count and after-pulsing will be performed elsewhere.

To characterise the efficiency we use a continuous DFB (distributed feedback) diode laser source at 1550\,nm with a calibrated attenuator so as to send the required average number of photons to the detector.  The bias voltage on the APD is varied and the counts corresponding to all detections are registered. To record the noise counts, we use an optical shutter to remove the laser signal. In the free-running scheme we determine the  {\it quantum} efficiency in the following way:
\begin{equation}
\eta_Q = \left [ S/(1- S\, \tau_{d}) - N/(1- N\,\tau_{d}) \right ]\,/\,n
\end{equation}
where we have corrected for the dead-time $\tau_{d}$ of the detector. We denote the signal $S$, detected when $n$ photons (per second) are sent to the detector, and the noise is given by $N$. This gives the probability of detecting a photon that is incident on the detector while it is active (this is, by definition, the case for photon counting in the gated regime). However, what we are really interested in is the {\it Effective} efficiency, where we place no constraint on the detector being ready for a photon that is there, as should be the case for a free-running detector. To calculate this, we make no corrections for the dead-time and define $\eta_{Eff} = (S - N)/n$. This efficiency will saturate as more detections force the detector to be active less often.

\begin{figure}
\begin{center}
\epsfig{figure=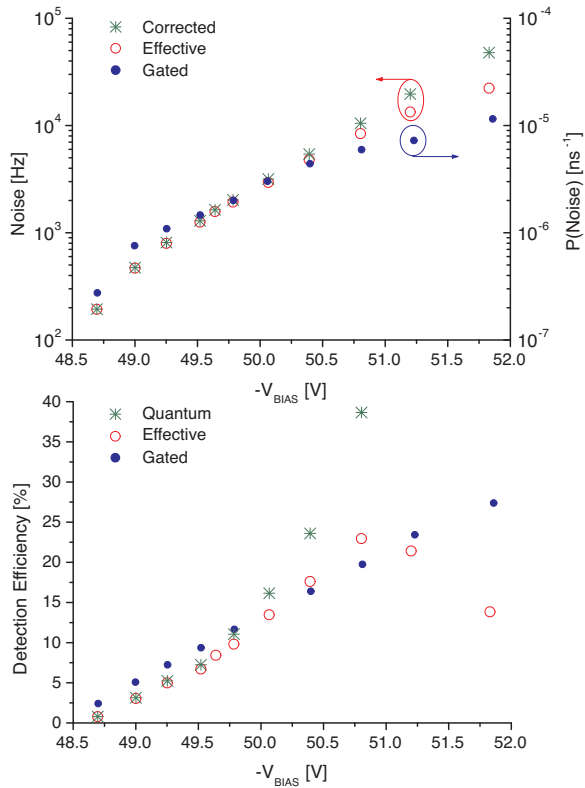,width=80mm}
\caption{Free-running and gated mode, noise and efficiency characteristics, as a function of APD bias voltage.}
\label{fig:EffVsNoise}
\end{center}
\end{figure}

We have compared the operation of a standard InGaAs/InP APD (JDSU-EPM657SS) combined with the ASIC chip operating in both gated mode, with a gate width of 100\,ns, and free running with our external control circuit. In the case of Fig.\,\ref{fig:EffVsNoise}, we see the results for both: free running, using $n = 10^4$ photons s$^{-1}$ and a dead-time of 24\,$\mu$s; and gated, with 1 photon per pulse at a trigger frequency of 10\,kHz (the same number of photons in each case),  and a dead-time of 200\,$\mu$s. The noise characteristics are shown at the top of Fig.\,\ref{fig:EffVsNoise}, where there is good agreement between the two cases. There is a small variation at higher bias voltages (V$_{BIAS}$) between the effective (measured) and corrected (for dead-time) noise in the free-running case. The gated mode response has a slightly different slope but the general behaviour is comparable. In the lower graphic we again see that the gated and free-running efficiencies differ in a similar manner to the noise. Importantly, we note that while  the quantum efficiency increases to more than 30\,\%, the {\it Effective} efficiency saturates depending on the dead-time. In general we see good agreement for moderate efficiencies around 10\,\%. The slight variations in the noise and efficiency characteristics, as well as the after-pulsing, are currently under further investigation.


In conclusion, we have presented the results for a free-running InGaAs/InP single photon detector working with an active quenching ASIC. The rapid quenching provided by the ASIC chip significantly reduces the problems with after-pulsing that have previously rendered this mode of operation unattainable. We obtain a detection efficiency of 10\,\% with $<$\,2kHz of noise at 1550\,nm with an after-pulse probability $<$\,8\,\% for $\tau_{d} =$ 24\,$\mu$s. At  $\tau_{d} =$ 32\,$\mu$s this after-pulse probability has reduced to less than 1\,\%. No discernible difference in the timing jitter ($<$\,400\,ps at 10\,\% detection efficiency) between gated and free running modes is observed. Our detector is simple to operate with continuous, or asynchronous, sources and has a distinct advantage in the case where one is trying to detect low numbers of photons.  In this regime, one must take account of the time that the detector is active, i.e. in the gated system this is given by $f_{Trig}$ x gate-width. In this sense, the free running detector has a significant advantage due to its higher active-detection duty-cycle. At high count rates, gated detection may still have an advantage, however, if after-pulsing is not so critical, a smaller dead-time could be used, also allowing higher counting rates in the free-running mode. 

The authors acknowledge useful discussions with Y.~Leblebici and C.~Guillaume-Gentil, as well as financial support from the Swiss NCCR Quantum Photonics and the Swiss CTI.

\end{document}